# Dynamicasome—a molecular dynamics-guided and AI-driven pathogenicity prediction catalogue for all genetic mutations

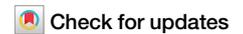 Check for updates

Naeyma N. Islam[1], Mathew A. Coban [1], Jessica M. Fuller [1], Caleb Weber[1], Rohit Chitale [2], Benjamin Jussila [3], Trisha J. Brock[3], Cui Tao[4] & Thomas R. Caulfield [1,4,5,6,7,8,9,10] ✉

Advances in genomic medicine accelerate the identification of mutations in disease-associated genes, but the pathogenicity of many mutations remains unknown, hindering their use in diagnostics and clinical decision-making. Predictive AI models are generated to combat this issue, but current tools display low accuracy when tested against functionally validated datasets. We show that integrating detailed conformational data extracted from molecular dynamics simulations (MDS) into advanced AI-based models increases their predictive power. We carry out an exhaustive mutational analysis of the disease gene PMM2 and subject structural models of each variant to MDS. AI models trained on this dataset outperform existing tools when predicting the known pathogenicity of mutations. Our best performing model, a neuronal networks model, also predicts the pathogenicity of several PMM2 mutations currently considered of unknown significance. We believe this model helps alleviate the burden of unknown variants in genomic medicine.

A major obstacle to achieving higher diagnostic rates in genomic medicine is the large number of variants of uncertain significance (VUSs). With currently available data, it is often challenging to definitively determine whether a variant is pathogenic or benign, even when identified in a gene strongly associated with disease. This ambiguity often results in mutations remaining non-diagnostic and unsuitable for clinical decision-making[1], complicating efforts to identify and treat genetic disorders[2]. The advent of inexpensive sequencing technologies and the integration of global genomic information have exponentially increased the number of VUSs in databases like ClinVar[3], and misclassification of these variants poses a critical concern that could lead to severe consequences for patient care[4]. Thus, there is an urgent need for more accurate and efficient methods to distinguish between pathogenic and non-pathogenic variants. This is no small feat: there are approximately 20,000 genes constituting the human genome, and OMIM (omim.org) reports that over 5000 of these genes are currently associated with disease or disease risk[5]. The potential for genetic variation is therefore immense, and directly testing the pathogenicity of variants at such a large scale is unfeasible. Computational approaches capable of predicting the pathogenicity of mutations have therefore emerged as a viable means of combating the burden of VUSs in genomic databases[6,7].

Diverse predictive software tools have been developed to categorize VUSs, and current guidelines recommend a concordance-based approach using multiple prediction algorithms[1]. Unfortunately, there is no clear guidance on which tools to use or how to achieve consensus, leading to inconsistencies across clinical laboratories[8]. Meta-predictors like rare exome variant ensemble learner (REVEL), which integrate information from multiple sources into machine-learning algorithms, have been used to address these inconsistencies[9–11], but have unfortunately had limited success when tested against clinically relevant and functionally validated datasets[12,13]. Concordance approaches using these tools also suffer from the same drawbacks as the individual algorithms, potentially leading to false positives when similar tools result in concurrent misclassifications[8,12]. The suboptimal performance of these tools may be the specific gap in the data related to missing protein-dynamic information they use as inputs: most predictive methods employed to date, including those incorporated into meta-predictors, focus on sequence-derived data and measures of evolutionary conservation to predict the pathogenicity of mutations. This includes alignment-based scores like those generated by tools like Protein Variation Effect Analyzer (PROVEAN)[14,15] and data garnered from the analysis of 3D structural models using tools like RFold. Studies suggest that both AlphaFold and conservation-only models show low accuracies when tested against functional assay-derived datasets[12,16], suggesting that these approaches do not incorporate detailed structural dynamics, which limits their ability to predict the nuanced impacts of mutations on protein function.

[1]Department of Neuroscience, Mayo Clinic, Jacksonville, FL, USA. [2]Department of Infectious Disease, Mayo Clinic, Jacksonville, FL, USA. [3]InVivo Biosystems, Inc., Eugene, OR, USA. [4]Department of Artificial Intelligence and Informatics, Mayo Clinic, Jacksonville, FL, USA. [5]Quantitative Health Sciences, Biostatistics Division, Mayo Clinic, Jacksonville, FL, USA. [6]Department of Biochemistry and Molecular Biology, Mayo Clinic, Rochester, MN, USA. [7]Department of Neurosurgery, Mayo Clinic, Jacksonville, FL, USA. [8]Department of Cancer Biology, Mayo Clinic, Jacksonville, FL, USA. [9]Department of Clinical Genomics, Mayo Clinic, Rochester, MN, USA. [10]Digital Ether Systems and Computing, Inc, Miami, FL, USA. ✉e-mail: thomas@digitalethercomputing.com

The role of dynamics and structural changes in proteins is crucial for understanding the impact of mutations, highlighting the need for improved predictive models[17–19]. Structural dynamics influence how proteins fold, interact with other molecules, and perform their biological functions. Mutations can alter these dynamics, leading to changes in protein stability, flexibility, and function. Studies have shown that incorporating dynamic features, such as those obtained from molecular dynamics simulations (MDS), can significantly enhance the accuracy of mutation impact predictions[20]. For instance, MDS provide insights into conformational changes, stability, and interactions at an atomic level, which are critical for assessing the functional consequences of mutations[21–29]. Building on this understanding, our study addresses the shortcomings of current predictive tools for missense variants by integrating MDS and advanced artificial intelligence (AI). Unlike traditional methods, which often rely on sequence-based features and evolutionary conservation data[30,31], our approach incorporates detailed structural and dynamic insights from MDS. This enables a comprehensive assessment of protein mutation impacts, allowing for a more accurate prediction of the pathogenicity of variants. By combining these sophisticated computational techniques, our goal is to significantly reduce the number of Variants of Uncertain Significance (VUS) and improve the accuracy of distinguishing pathogenic from benign mutations. Our approach takes exhaustive examination of all possible missense mutants of a protein and completes simulations for the dynamic motion that allows the propagation of the missense variant to occur (Schematic S1), which allows the genetic mutation to perturb the protein, revealing its pathogenic capacity. These dynamic results can be correlated to clinical and biochemical data in a robust AI modeling system for predictions.

To develop and test our approach, we focused on phosphomannomutase 2 (PMM2), a small enzyme (28 kDa) (Fig. 1a) responsible for the isomerization of mannose-6-phosphate to mannose-1-phosphate. Pathogenic variants of *PMM2* cause congenital disorder of glycosylation type Ia, also known as PMM2-CDG: a genetic disease characterized by encephalopathy, movement disorders, and cerebellar hyperplasia[32]. Over 100 mutations in PMM2 have been associated with PMM2-CDG, and each may exert varying effects on PMM2 stability and enzymatic function[33,34]. Unfortunately, despite its clinical significance, the precise pathogenicity of many PMM2 missense mutations remains unclassified. We conducted an exhaustive mutational analysis of PMM2 by single-nucleotide polymorphisms in the cDNA, which, when translated into the corresponding protein, allows for sampling all possible variations for missense variants in the gene. In the case of single substitutions, we find there are around 1454 variants for the gene encoding PMM2 (NCBI NP_000294, Uniprot accession #O15305) that occur from the 20 standard amino acids. We then generated 3D structures for each variant and subjected them to MDS. By extracting multiple key features from the MDS data, we provide a comprehensive view of how each individual mutation impacts the conformational dynamics and stability of the protein. These features were then used to train various advanced AI models, the performance of which was measured against existing approaches like REVEL, PROVEAN and AlphaMissense. Our models showed superior performance compared to existing benchmark models, arguing that their ability to capture the intricate dynamic and structural changes caused by missense mutations represents a crucial step forward in predictive modeling.

To further validate the predictions generated by our model, we tested a subset of PMM2 variants using an in vivo system, the *C. elegans* nematode humanized with PMM2. Humanization enables studies in a whole animal using the human protein, which provides a physiologically relevant context for analyzing variant effects[35–37]. Our model outperformed existing tools such as REVEL in predicting variant status for the four variants tested. We believe our approach will not only advance our understanding of the role of PMM2 variants in PMM2-CDG but also improve the accuracy of pathogenicity predictions and reduce the number of VUSs in clinical databases.

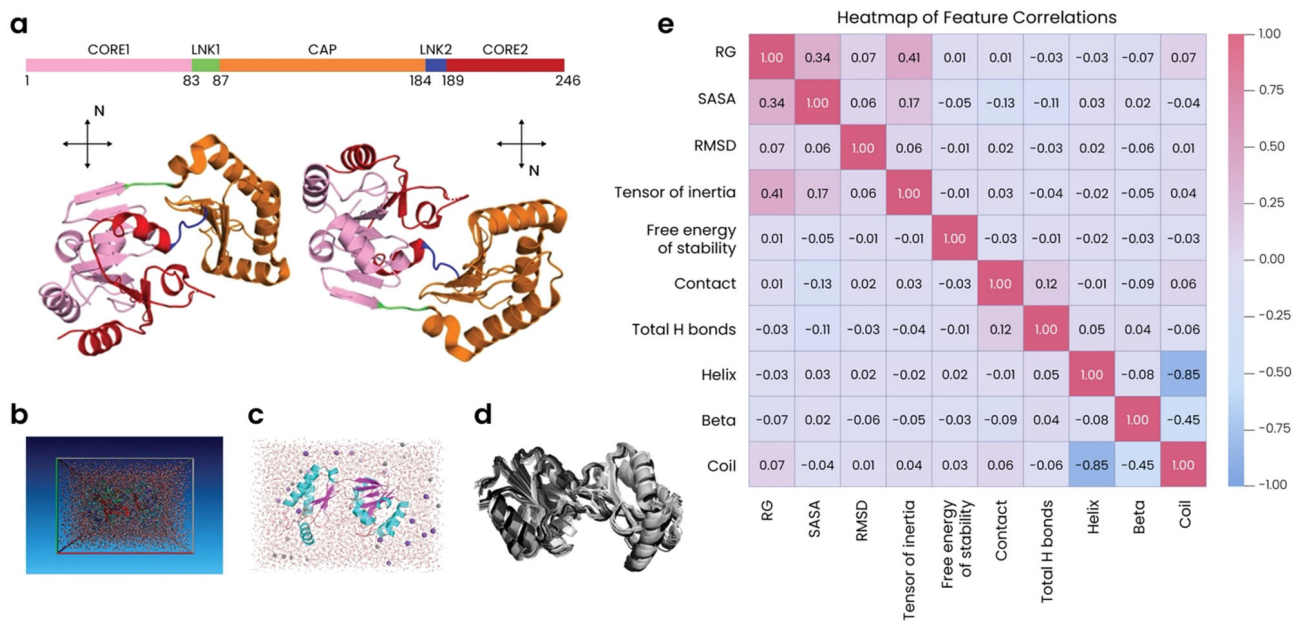

**Fig. 1 | Features extracted from MDS of PMM2 variants show a range of correlations. a** Structural map of wildtype PMM2 shown in two rotations. Domains are color-coded according to the schematic above and include a CAP domain involved in dimerization and activator binding (orange) and two CORE domains (pink and red) involved in binding catalytic and structural $Mg^{2+}$ ions. Two linker (LNK) domains (green and blue) provide flexibility, which enables motions necessary for catalysis. **b,c** MDS were run in a closed system (**b**) in a simulated physiological environment (**c**). Water molecules are depicted as red and white sticks, $Na^+$ ions as gray spheres, and $Cl^-$ ions as purple spheres. PMM2 is color-coded by secondary structure, with α-helices in blue, β-sheets in magenta, and loops or coils in salmon. **d** 20 conformations of wildtype PMM2 overlaid upon each other demonstrate dynamics throughout the simulation. Gray darkness increases as the simulation progresses. **e** Heatmap of correlations between MDS features. The color scale ranges from −1.00 (blue), indicating a perfect negative correlation, to 1.00 (red), indicating a perfect positive correlation.

## Results

### MDS reveals the impact of missense mutations on PMM2 conformational dynamics

We analyzed the sequence of PMM2 to determine all possible missense mutations and then used in silico substitution to introduce each mutation into the PMM2 structure, resulting in 1454 different variant models. We then embedded each model into a physiological environment and ran an identical MDS protocol (Fig. 1b–d), allowing us to observe how each mutation affects the protein's intrinsic motions over time compared to the dynamics of wild-type (WT) PMM2 (Fig. 1a). Following each MDS, we extracted multiple quantifiable properties we refer to as "features," including the radius of gyration (Rg), the degree of solvent-accessible surface area (SASA), the root-mean square deviation (RMSD), the tensor of inertia, the free energy of stability, the amount of amino acid contacts, the number of hydrogen bonds, and the percentage of the protein that is arranged into secondary structure elements like α-helices, β-sheets, and coils. All features were standardized to ensure meaningful comparisons; each feature varied in response to different mutations, with some showing a wider distribution across mutations than others (Fig. S1).

We then visualized the correlations between various features using heatmaps (Fig. 1e) and pair plots (Fig. S2) to determine whether any features had interconnected roles in protein structure and stability. These analyses revealed that some features, like Rg and the tensor of inertia, showed some positive correlation, while other features, like RMSD and SASA, showed only low correlations that suggest they are likely influenced independently (Fig. 1e).

### MDS features show more variability in response to pathogenic PMM2 mutations than benign mutations

Having determined the effect of each mutation on the structural dynamics of PMM2, our next step was to investigate the relationship between these features and disease pathogenesis. We meticulously labeled the mutations by their predicted pathogenicity by evaluating clinical genetic sequencing results reported in ClinVar[38–40] (Table S1) and the Human Gene Mutation Database[41]. We also mined the genome aggregation database (gnomAD)[42,43] for PMM2 mutations; as this database only contains data from individuals that are 18 years or older, and PMM2-CDG presents at a very early age with an autosomal recessive inheritance pattern, we concluded that all homozygous PMM2 variants detected in this database are benign, even if they are classified as likely benign or VUS in ClinVar (Table S2). Finally, we collaborated with the Center for Individualized Medicine (CIM) at Mayo Clinic via the clinical genomics department, which had identified patients carrying compound heterozygous pathogenic PMM2 variants, to gain additional clinical insight into damaging and benign mutations. In total, we were able to classify 54 mutations as benign, 97 mutations as damaging, and 6 mutations as ambiguous. Here, we use the terms "VUS" and "ambiguous" interchangeably to refer to the same category of mutations. The remaining 1297 mutations are of unknown pathogenicity.

Focusing on the subset of mutations that we were able to label, we used violin plots to visualize the relationship between MDS-extracted features and pathogenicity (Fig. 2a–j). Overall, features tended to show more narrow distributions in benign mutations—often centered around a median close to zero, indicating minimal variation—and wider distributions in damaging mutations (Fig. 2a–j). The widest distributions were observed for ambiguous mutations across all features (Fig. 2a–j).

### Advanced AI models can classify variants by their pathogenicity and outperform existing predictive models

Bolstered by our observation that MDS features show more variability in response to pathogenic PMM2 mutations than benign ones, we suspected that incorporating MDS features would improve the performance of predictive models aimed at classifying VUSs. Our dataset of labeled PMM2 mutations is unfortunately imbalanced, with substantially more damaging mutations than benign or ambiguous variants. Such datasets are not ideal for training AI models, and we therefore tested multiple methods of balancing our dataset, including oversampling the minority classes, downsampling the majority class, the synthetic minority oversampling technique (SMOTE)[44], algorithmic adjustments, and ensemble methods like bagging and boosting. SMOTE proved to be the most effective approach. We then used our SMOTE-balanced dataset to train multiple AI-based classification models via a stratified 10-fold cross-validation approach. The dataset was split into 10 subsets, and each model was trained on 9 of these sets before being validated on the tenth. This process was then repeated so that each subset served as the test set once; the results were averaged. We tested 7 AI models based on distinct machine-learning methods: random forest (RF), deep neural networks (DNN), semi-supervised learning (SSL), decision tree (DT), logistic regression (LR), gradient boosting classifier (GBC), support vector machine with a radial basis function kernel (SVM-rbf), and k-nearest neighbors (KNN). The performance of each model was assessed using receiver operating characteristic (ROC) curves and their corresponding area under the curve (AUC) values. We applied a one-vs-rest (one-vs-all) strategy when computing ROC curves for each class individually, which is a standard approach for adapting ROC analysis to multiclass problems. For comparison, we also performed the same analysis with established benchmark models, namely REVEL[9], PROVEAN[14,15], and AlphaMissense, a recent tool that uses AlphaFold technology fine-tuned on human and primate variant datasets to allow for the incorporation of both structural context and evolutionary conservation into prediction readouts[30,45]. In future, to strengthen our variant classification approach for PMM2 mutations, we aligned with recent ClinGen guidelines recommending the use of calibrated computational tools for PP3/BP4 criteria (Bergquist et al., 2024)

For benign mutations, the SSL model achieved the highest AUC of 0.92, effectively leveraging both labeled and unlabeled data (Fig. 3a). The RF, DNN, KNN, and GBC models also performed well, while the SVM-rbf and DT models showed moderate performance (Fig. 3a). All advanced AI models outperformed the benchmark models and logistic regression, the latter of which achieved an AUC akin to random guessing (Fig. 3a).

Similar results were observed for damaging mutations, where the DNN model achieved the highest AUC of 0.87, and the semi-supervised learning, RF, and GBC models all achieved an AUC of 0.76 (Fig. 3b). The SVM-rbf and DT models once again showed moderate performance (Fig. 3b). The KNN model performed worse with this dataset compared to its performance with benign mutations, achieving an AUC that was comparable to that of the benchmark models—all of which performed slightly better when classifying damaging mutations compared to benign (Fig. 3b). LR once again achieved the lowest AUC (Fig. 3b).

Finally, for ambiguous mutations, the DNN model achieved an AUC of 1.00, indicating perfect classification (Fig. 3c). The other advanced AI models also achieved high AUC values, with the KNN model showing the worst performance out of the 7 models. All our models again outperformed the benchmark methods, which achieved AUCs comparable to that of random chance (Fig. 3c). Interestingly, LR proved moderately adept at classifying ambiguous mutations, achieving an AUC of 0.77. Its performance was still less robust than that of the advanced AI models, however (Fig. 3c).

### The DNN model showed the best performance by averaged ROC-AUC across multiple mutation classes

When we averaged the ROC values across all mutation classes, the DNN model achieved the highest average AUC of 0.90 (Fig. 3d), showcasing its ability to capture complex patterns. The RF and SSL models also performed well (Fig. 3d), leveraging ensemble methods and both labeled and unlabeled data, respectively. The GBC and SVM-rbf models had moderate average AUCs (Fig. 3d), benefiting from boosting and non-linear handling. The KNN model showed moderate performance, while the AUC of the DT model was slightly lower due to overfitting (Fig. 3d). Logistic regression, AlphaMissense, REVEL, and PROVEAN had the lowest average AUCs (Fig. 3d), limited by their inability to manage high-dimensional and imbalanced data.

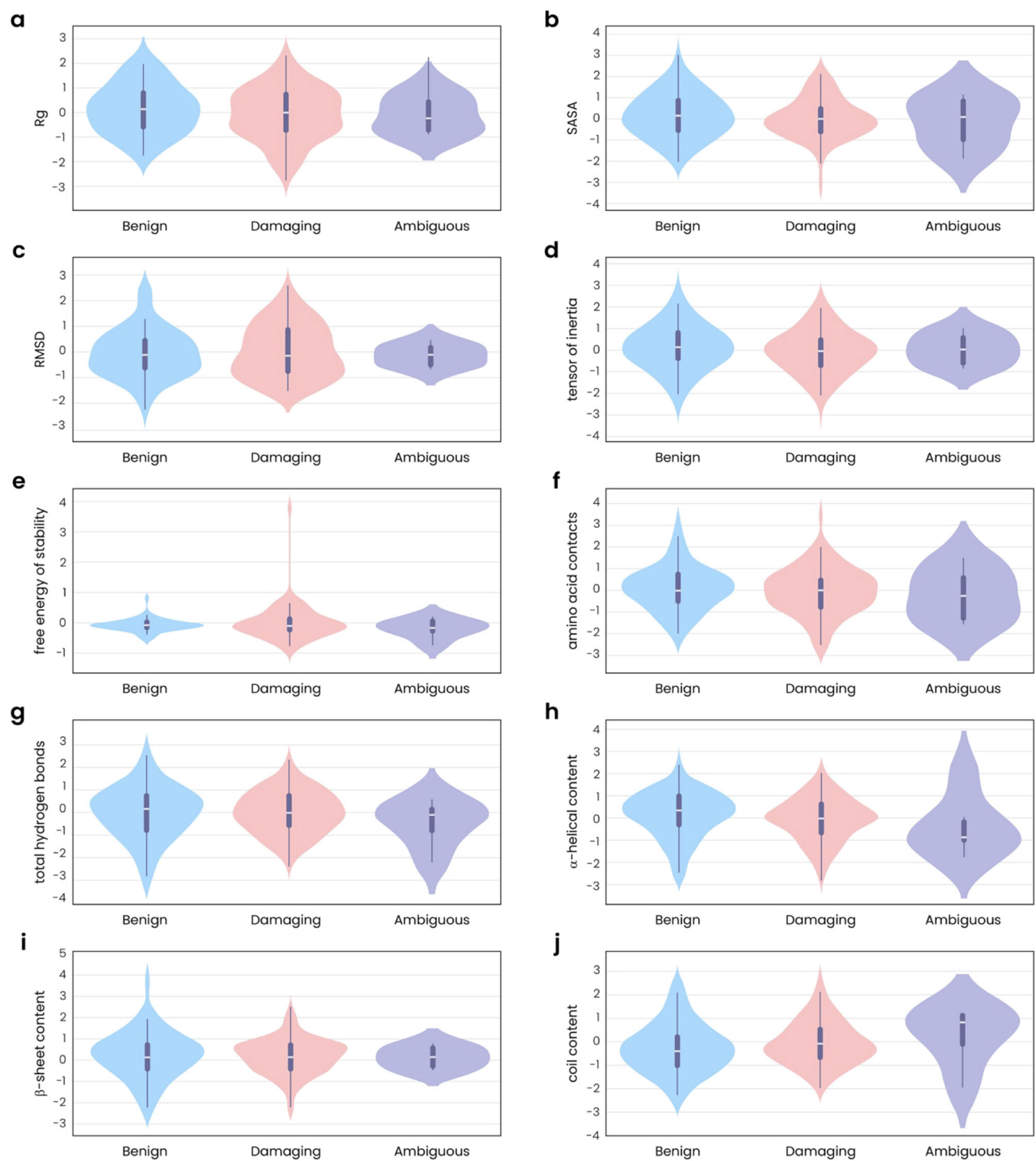

**Fig. 2 | MDS features show more narrow distributions with benign mutations and increased variability with damaging and ambiguous mutations.** Violin plots showing the distribution patterns of **a** Rg, **b** SASA, **c** RMSD, **d** the tensor of inertia, **e** the free energy of stability, **f** the number of amino acids within 0.5 Å of one another, **g** the total number of hydrogen bonds, **h** the percent of α-helical content, **i** the percent of β-sheet content, and **j** the percent of coil content in datasets of benign (blue), damaging (pink), and ambiguous (purple) PMM2 mutations. All feature values were scaled for comparison.

## The DNN and RF models showed the highest overall predictive accuracy

In addition to ROC-AUC, we also quantified the average accuracy of each model across all mutation classes as a second readout of overall performance (Fig. 3e). The RF and DNN models showed the highest accuracy, indicating strong generalization capabilities (Fig. 3e). The SSL model also performed well, while the other advanced AI models showed moderate overall accuracy. LR exhibited the lowest accuracy (Fig. 3e), highlighting its limitations in handling non-linear data. Interestingly, the existing tools (particularly REVEL and PROVEAN) show relatively high accuracy (Fig. 3e), contrasting their lower ROC-AUC values (Fig. 3d).

We used confusion matrices to gain additional insights into the ability of the models to classify different PMM2 mutations. Both the RF (Fig. 4a) and SSL (Fig. 4b) models showed exceptional accuracy with high true positive rates for benign and ambiguous mutations and only minor misclassifications of damaging class instances. The DNN (Fig. 4c) and GBC (Fig. 4d) models also maintained robust classification capabilities with some misclassification of damaging mutations. The KNN (Fig. 4e), SVM-rbf (Fig. 4f), and DT (Fig. 4g) models showed moderate accuracy, although a notable number of damaging class instances were misclassified as benign with the DT method (Fig. 4g). These models all outperformed existing approaches: LR exhibited significant misclassifications (Fig. 4h), while the benchmark tools, although capable of accurately classifying a large number of damaging mutations, called little to no true positive benign or ambiguous instances (Fig. 4i–k).

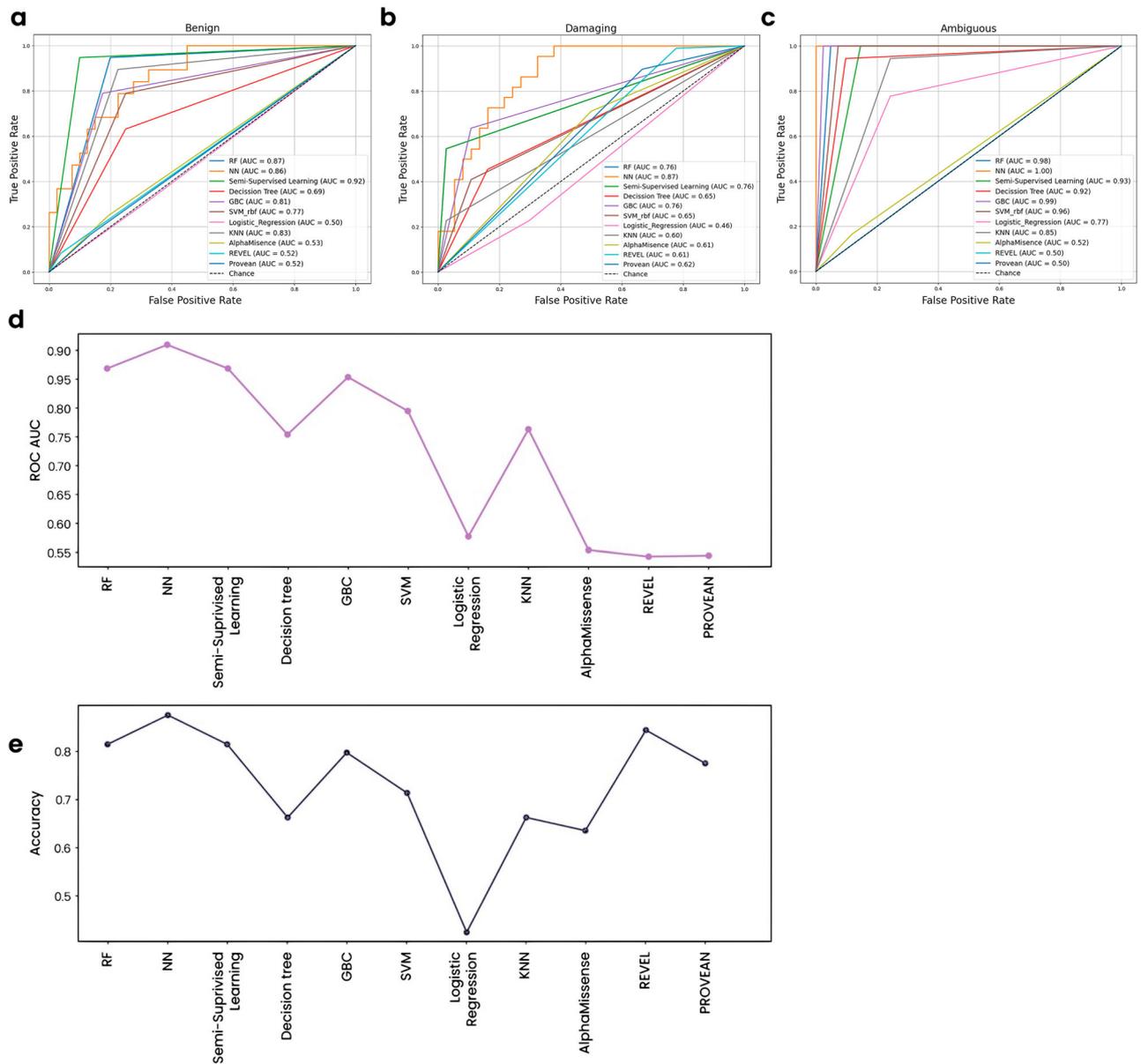

**Fig. 3 | Advanced AI models outperform benchmark and traditional approaches when classifying benign, damaging, and ambiguous PMM2 mutations. a** ROC plots for each model when classifying benign PMM2 mutations. AUCs (highest to lowest): SSL: 0.92, RF: 0.87, DNN: 0.86, KNN: 0.83, GBC: 0.81, SVM-rbf: 0.77, DT: 0.69, AlphaMissense: 0.53, REVEL: 0.52, PROVEAN: 0.52, logistic regression: 0.50. **b** ROC plots for each model when classifying damaging PMM2 mutations. AUCs (highest to lowest): DNN: 0.87, RF: 0.76, semi-supervised learning: 0.76, GBC: 0.76, SVM-rbf: 0.65, DT: 0.65, PROVEAN: 0.62, AlphaMissense: 0.61, REVEL: 0.61, KNN: 0.60, logistic regression: 0.46. **c** ROC plots for each model when classifying ambiguous PMM2 mutations. AUCs (highest to lowest): DNN: 1.00, GBC: 0.99, RF: 0.98, SVM-rbf: 0.96, semi-supervised learning: 0.93, DT: 0.92, KNN: 0.85, logistic regression: 0.77, AlphaMissense: 0.52, REVEL: 0.50, PROVEAN: 0.50. **d** Average ROC-AUC values across all three mutation classes for each model. **e** Accuracy of each model across all three mutation classes.

### Model performance was evaluated using F1 score, precision, recall, macro-F1, and micro-F1 metrics

Among the tested models, Random Forest (RF) achieved the best overall performance (F1: 0.804, macro-F1: 0.811, micro-F1: 0.814), followed closely by Semi-Supervised Learning (SSL) (F1: 0.803, recall: 0.836). Both Gradient Boosting Classifier (GBC) and Deep Neural Network (DNN) showed strong results, with macro-F1 scores above 0.73. In contrast, Logistic Regression (F1: 0.413) and KNN (F1: 0.629) underperformed, indicating limited classification accuracy for these models. Complete performance metrics are provided in Table 1.

### RMSD is the most important feature for predictive modeling with MDS data

We used a DT classifier to determine the relative importance of each MDS-extracted feature to the predictive ability of our AI-based models. RMSD emerged as the most critical feature, with the highest importance at approximately 40% (Fig. 5). Rg, which provides insight into the protein's compactness, is also highly important, while the free energy of stability ranks

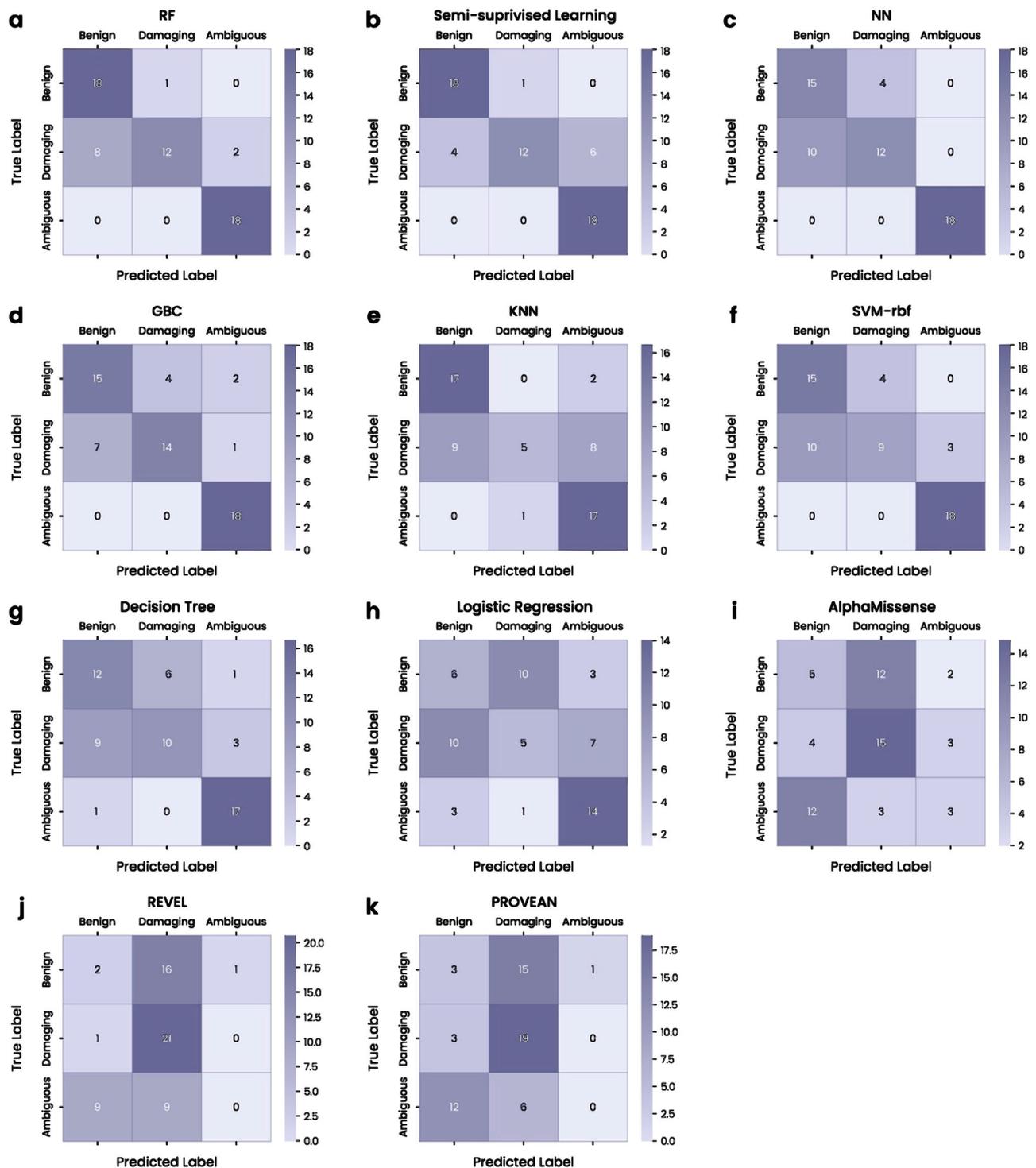

**Fig. 4 | Confusion matrices highlight the relative success of different models when classifying different classes of PMM2 mutations.** The number of mutations predicted to be of each class (benign, damaging, or ambiguous) is plotted against the true labels, so that each matrix shows the number of true positive, false positive, false negative, and true negative predictions for the advanced AI models: **a** RF, **b** SSL, **c** DNN, **d** GBC, **e** KNN, **f** SVM-rbf, and **g** DT. Matrices are also shown for LR (**h**) and the three benchmark models, **i** AlphaMissense, **j** REVEL, and **k** PROVEAN. Darker shades indicate higher counts.

third (Fig. 5). The coil content, while still relevant, showed the lowest importance (Fig. 5).

### Different AI models show slightly different predictions for the classification of VUSs

Our ultimate goal is to develop predictive models capable of accurately classifying VUSs as either pathogenic or benign, as such tools would be of supreme benefit in the clinic. We therefore used our trained AI models and the existing benchmark techniques to independently label an identical set of PMM2 mutations currently classified as VUSs. We visualized this data as binary heatmaps (Fig. S3) and as bar graphs (Fig. 6a). Each model's predictions varied, demonstrating their individual capabilities in categorizing mutations into the respective categories (Fig. 6a). The DNN model predicted the highest percentage of benign mutations (59.0%), while REVEL predicted the highest percentage of damaging mutations (81.6%) (Fig. 6a). The SSL model showed a notable percentage of ambiguous predictions (22.2%).

We performed a comparative analysis of the predictions made by our best-performing model, the DNN model, against AlphaMissense, REVEL, and PROVEAN (Fig. 6b). Out of 1297 PMM2 mutations considered VUSs, the DNN model predicted 478 benign mutations—105 of which were also predicted by AlphaMissense, REVEL, and PROVEAN (Fig. 6b). It also identified 258 damaging mutations, 216 of which were consistent with other programs (Fig. 6b).

### In vivo confirmation of AI model predictions

To validate the predictions generated by our AI model, we utilized *C. elegans*, which possesses a homolog of the human PMM2 gene, known as F52B11.2[46]. Knock-out of F52B11.2 in *C. elegans* results in a lethal, sterile phenotype, making it an effective proxy for studying PMM2 function[47]. By replacing F52B11.2 with a codon-optimized human PMM2 transgene, we successfully humanized the system and demonstrated functional rescue of the lethal phenotype. This humanized strain provides a robust platform for assessing the functional consequences of clinical variants in PMM2.

We tested a subset of PMM2 variants, focusing on those with pathogenicity predictions from our AI model that conflicted with benchmark tools like REVEL. Variants were introduced into the humanized *C. elegans* strain via CRISPR-Cas9 genome editing, and their impact was assessed by evaluating the presence or absence of the lethal phenotype.

The p.Q37R variant was predicted to be benign by both our AI model and REVEL, and it did not result in a lethal phenotype, supporting the predictions of both models (Table 2). In contrast, the p.C241Y variant was predicted to be benign by our model but pathogenic by REVEL. The animals did not exhibit a lethal phenotype, aligning with the prediction of our AI model. The p.R141L variant, predicted to be pathogenic by both our model and REVEL, displayed a lethal, sterile phenotype, confirming its deleterious impact on PMM2 function. The p.A19S variant, predicted to be benign by REVEL but pathogenic with our model, showed an intermediate phenotype. The variant was homozygously viable, but observations suggest that they produce less progeny. We are working to quantify this phenotype, but it suggests that the variant protein is not fully functional, which is consistent with the prediction of our AI model.

### Discussion

The classification of VUSs remains a critical problem that hinders disease diagnostics and research. AI-based predictive models represent a viable way of combatting this issue, but current models remain less than ideal—potentially due to their inability to take detailed conformational changes due to mutation into account. In this study, we conducted an exhaustive mutational analysis of PMM2, followed by the construction of 3D structures for each PMM2 protein variant and subsequent MDS to observe mutation-dependent conformational changes. By extracting key properties from the MDS, we aimed to provide a comprehensive view of PMM2 mutation impacts. We then used this information to train multiple AI-based models and measure their predictive efficacy in comparison to established

**Table 1 | Performance comparison of various machine-learning models in multi-class mutation prediction, evaluated using F1 score, precision, and recall**

| Models | F1 Score | Precision | Recall | macro-F1 | micro-F1 |
|---|---|---|---|---|---|
| DT | 0.655 | 0.655 | 0.661 | 0.6612 | 0.6610 |
| GBC | 0.778 | 0.778 | 0.779 | 0.7686 | 0.7627 |
| KNN | 0.629 | 0.713 | 0.661 | 0.6228 | 0.6610 |
| LR | 0.413 | 0.396 | 0.423 | 0.4152 | 0.4237 |
| RF | 0.804 | 0.841 | 0.813 | 0.8110 | 0.8136 |
| SSL | 0.803 | 0.813 | 0.836 | 0.8070 | 0.8136 |
| SVM | 0.700 | 0.712 | 0.711 | 0.7064 | 0.7119 |
| DNN | 0.774 | 0.801 | 0.779 | 0.7317 | 0.7288 |
| AlphaMissense | 0.682 | 0.660 | 0.793 | 0.3767 | 0.6609 |
| PROVEN | 0.769 | 0.773 | 0.754 | 0.3435 | 0.7739 |
| REVEL | 0.853 | 0.869 | 0.861 | 0.6377 | 0.8696 |

RF and SSL exhibit strong, balanced performance, while LR and KNN show weaker results. Benchmark models REVEL, PROVEN, and Alphamissense are also included for comparison.

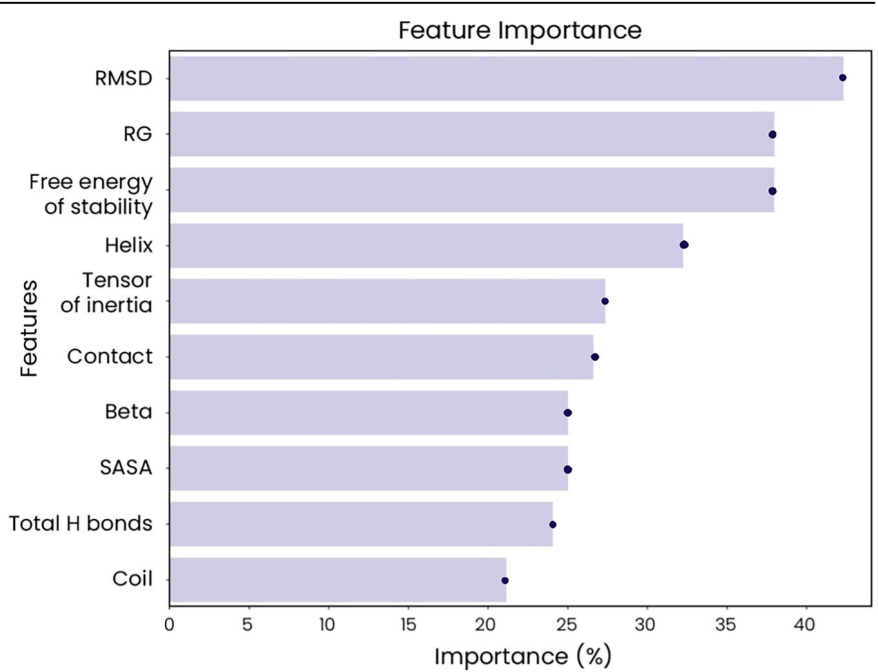

**Fig. 5 | Ranking of features by their contribution to the performance of advanced AI models reveals that RMSD is the most important factor when predicting PMM2 mutations.** The importance of each feature to the performance of our advanced AI models is plotted by percentage.

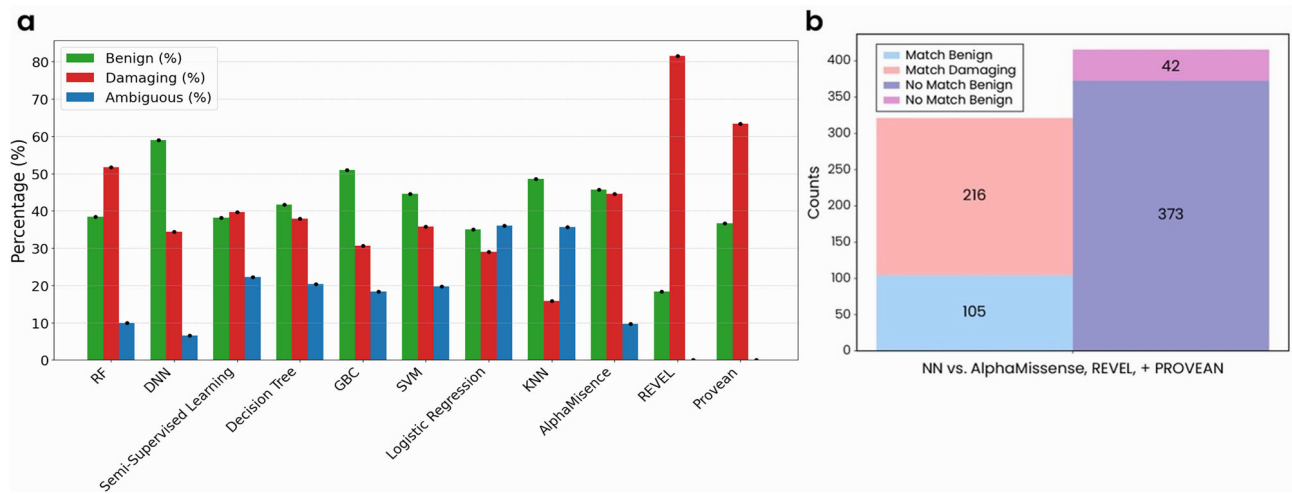

**Fig. 6 | Predictive models show varying outcomes when classifying PMM2 mutations of unknown significance. a** The bar graph illustrates the percentage distribution of predictions for each category (benign in blue, damaging in orange, ambiguous in green) across the indicated models. **b** Comparative analysis of DNN model predictions versus the predictions of benchmark models AlphaMissense, REVEL, and PROVEAN. When a mutation was predicted as benign or damaging by the DNN model and at least one benchmark model, they were categorized as a match: matched benign mutations are in blue and matched damaging mutations in orange. Predictions made by the DNN model that were not called by any benchmark models are labeled as "No Match Benign" in green and "No Match Damaging" in red.

benchmark approaches. Our models outperformed this established program, confirming the advantages of integrating MDS data into mutation classification models. Other laboratories have undertaken the utilization of MDS as a predictor of pathogenicity[48,49]; however, we present an integrated, exhaustive set of all missense variants (non-truncating) using dynamics simulations within an AI framework for predicting dysfunction that is biologically validated.

We extracted multiple quantifiable features from our MDS data and found that some features correlated, indicating their interconnected roles in protein structure and stability (Fig. 1e). For example, the moderate positive correlation between the Rg and the tensor of inertia suggests that more compact structures (i.e., lower Rg) are associated with different mass distributions, while the weak correlation between RMSD and SASA indicates that overall structural deviations have minimal effect on solvent exposure. The free energy of stability showed only weak correlations with other features, implying that stability impacts are generally independent of compactness, structural deviations, and specific structural elements. Overall, the relationships between these MDS-extracted features tell us about the intricate balance within the protein structure and shed insight into how the disruption of this balance upon mutation could impact protein function and contribute to disease pathogenesis. Accordingly, we found that many of these features show more variability in the presence of mutations classified as damaging compared to mutations labeled as benign (Fig. 2). More specifically, structures bearing benign mutations showed only minimal variations from the WT reference model (as indicated by the narrow distribution of RMSD values) with consistent secondary structures and stable internal interactions based on measurements of hydrogen bonds and amino acid contacts. These structures also tended to be more compact (lower Rg) with little changes in mass distribution (indicated by the tensor of inertia) and a consistent degree of solvent exposure. In contrast, structures bearing damaging mutations showed more expanded conformations that varied more widely from the WT structure, including a higher variability in amino acid contacts and secondary structure content. In fact, changes in RMSD proved to be the most important feature contributing to the success of our predictive models, and this was underscored by the relative importance of α-helical content and amino acid contacts (Fig. 5). Structures bearing damaging mutations also varied in their tensor of inertia, suggesting impacts on the protein's rotational properties, and showed less consistency in the amount of solvent exposure, which could impact the protein's stability and function. Indeed, damaging mutations generally appear to destabilize the protein (based on the free energy of stability, which is higher and more broadly distributed in these structures), while benign mutations have minimal or even positive effects on protein stability (Fig. 2e). Interestingly, mutations classified as ambiguous showed the highest variability in all these features, highlighting the profound impact of these features on PMM2 structure and function and the difficulties in classifying the role of some mutations in disease pathogenesis.

After training AI models on our PMM2 MDS data, we used two methods to evaluate their performance: ROC-AUC and accuracy. Accuracy measures the proportion of correct predictions, providing a straightforward measure of model performance. ROC-AUC, on the other hand, evaluates the trade-off between sensitivity and specificity, offering a more nuanced view of the model's ability to distinguish between classes across different thresholds. Both measures are important: while high accuracy indicates that a model is generally making correct predictions, a high ROC-AUC suggests that the model is effective at differentiating between classes, even if the dataset is imbalanced. Notably, genomic datasets, like the PMM2 dataset we used herein, are often imbalanced, as classifying mutations based on clinical data alone is often difficult. We find that the use of SMOTE to balance datasets improves the performance of our models, and more importantly, we find that advanced AI models trained on MDS data are more capable of handling complex and imbalanced datasets than existing tools. While some of the benchmark methods successfully exhibited high accuracy scores, they ultimately displayed low ROC-AUC values. It is worth noting that we observed some variation in the success of our individual AI-based models across different mutation classes, however. For instance, the SSL model demonstrated exceptional performance when classifying benign mutations due to its ability to leverage labeled and unlabeled data, while ensemble methods like RF and GBC proved adept at classifying ambiguous mutations, indicating their strength in handling varied and complex datasets. Overall, the DNN model showed the best performance across all mutation classes, with both high accuracy and the highest average ROC-AUC value.

**Table 2 | In vivo testing results confirm AI model predictions for PMM2 variants**

| Variant | Dynamicasome prediction | REVEL score | Phenotype Observed |
|---------|------------------------|-------------|---------------------|
| Q37R    | Benign                 | 0.377       | None                |
| R141L   | Pathogenic             | 0.986       | Lethal              |
| C241Y   | Benign                 | 0.901       | None                |
| A19S    | Pathogenic             | 0.358       | TBD                 |

Our analysis, based on F1 score, precision, recall, macro-F1, and micro-F1, reveals that while REVEL achieved the highest overall F1 score (0.853) and micro-F1 (0.870), its lower macro-F1 (0.638) indicated performance imbalance across classes. AlphaMissense and PROVEN also exhibited relatively low macro-F1 scores (0.377 and 0.344, respectively), suggesting reduced robustness in multiclass settings. In contrast, our top-performing models—particularly RF, SSL, and GBC—demonstrated balanced and reliable classification across benign, damaging, and ambiguous mutation categories, underscoring their potential utility for functional interpretation of PMM2 variants (Table 1).

The apparent strengths and weakness of different models when classifying different mutation classes were also evident when we challenged each method to label a set of PMM2 mutations currently considered VUSs (Fig. 6a). With this dataset, the DNN and GBC models displayed a strong capability in identifying benign mutations, while the RF model and established models like REVEL and PROVEAN appear more robust at calling damaging mutations. Models like LR and KNN had a significant proportion of predictions in the ambiguous category, suggesting a higher level of uncertainty in their classifications. When comparing the outputs of our DNN model with established methods, we found that the DNN model replicated a notable number of predictions made by benchmark models, and further classified a substantial number of additional benign and damaging mutations that were not labeled by other methods (Fig. 6b). We believe this difference in prediction outcomes suggests that our DNN model may capture nuances that the benchmark models miss, and therefore has the potential to offer more precise and reliable classifications on account of its heightened sensitivity and specificity. Further wet lab experiments focused on establishing the pathogenicity of the mutations our model predicts to be damaging could provide concrete evidence of its predictive power and ensure its practical applicability in the disease context.

Our in vivo modeling using humanized *C. elegans* provided further validation for the AI model predictions. Of the variants tested, two were classified similarly by our AI model and REVEL, and both were confirmed in *C. elegans*. The p.Q37R variant was predicted as benign by both models, which was consistent with the observation that the animals did not exhibit a lethal phenotype. The p.R141L variant, classified as pathogenic by both models, displayed a lethal, sterile phenotype, confirming its deleterious nature. In contrast, the in vivo results for the two variants, where predictions between our AI model and REVEL diverged, supported our AI model. The p.C241Y variant, predicted as benign by our model but pathogenic by REVEL, showed no lethal phenotype, aligning with our AI model. Similarly, the p.A19S variant, classified as pathogenic by our model but benign by REVEL, exhibited an intermediate phenotype. This variant was homozygous viable but produced fewer progeny, indicating partial functionality and supporting our model's prediction.

At the same time, we cannot rule out the possibility that the benchmark models are simply not ideal for the classification of PMM2 mutations, specifically. The effectiveness of predictive AI models depends heavily on the quality and comprehensiveness of the training data and the optimization of the algorithms and scoring systems integrated into the model. The benchmark models may not be properly trained and optimized for the unique characteristics of PMM2 mutations, potentially leading to less accurate predictions. Future approaches will increase time sampling to improve features for AI models, which requires monumental resource allocation.

In conclusion, by integrating the feature that captures detailed insights into the conformational flexibility and dynamic behavior of proteins from MDS into advanced AI-based predictive models, our study overcomes the limitations of sequence- and conservation-based tools that fail to fully represent the complex dynamics and structural changes induced by mutations in PMM2, offering a more accurate and comprehensive model for classifying PMM2 mutations. We believe this approach will also be instrumental in classifying mutations in other disease genes and could serve to reduce the high burden of VUSs that plague the diagnosis and treatment of genetic disorders. This innovative methodology advances the field of computational biology and bioinformatics, offering a powerful tool for the analysis of protein mutations and their implications in human health. One can envision a star trek future that encompasses the idea of Fenyman's "wiggling and jiggling" biomolecules and incumbent interactions containing the vast permutations of genetic variation calculated and prepopulated, as a Dynamicasome (likely from using QM computers and AI), which will give predetermination of likelihood of diseasedness in a lookup database. In future applications, to support clinical variant classification for PMM2 mutations, our model could be adapted to align with ACMG/ClinGen guidelines that recommend calibrated computational tools under PP3/BP4 criteria[50]. The integration of our precision-tailored AI platform for molecular glue design with ligand- or context-conditioned generative frameworks—such as the Ligase-Conditioned Junction Tree Variational Autoencoder (LC-JT-VAE)—offers a compelling path toward rational, patient-specific therapeutics. This approach is particularly well-suited to applications like Dynamicasome, where targeting the unique genetic and structural features of mutant proteins is essential for therapeutic efficacy[51].

## Methods
### Data generation workflow

Using in-house scripts, we created a workflow to date patient gene (accession number) to go from cDNA sequence (NCBI) and generate all missense variants encoded for that cDNA, which in the case of PMM2 is 1454 variants. The cDNA to protein FASTA generator is then fed into our Schrodinger workflow to model all protein structures (e.g., PRIME or RosettaFold) and then through Protein Prep Wizard (Schrodinger) to get a starting structure, which is minimized using conjugate gradient for 2500 steps. The final structures are archived as PDBs and MAE files for use in a parallel Molecular Dynamics simulation (MDS) engine that has a lookup table for archiving and sorting all of the simulation trajectories as they are captured. After 50 ns of MDS, every trajectory is archived for statistical mechanics files needed and then analyzed for standard statistical mechanics properties, like root-mean-square (RMS) deviation, fluctuation, radius-of-gyration, solvent-accessible surface area, and many other properties described in the artificial intelligence section. All scripts are made available on GitHub.

### Mutation generation and 3D structure modeling

A custom script was used to determine all possible single-nucleotide polymorphisms that would result in a missense variant, filtering out splicing and truncating variants. We then utilized the empirical structure of PMM2 (2AMY) as a seed for generating models of all 1454 variants.

### Molecular dynamics simulations (MDS)

We performed an identical 10 ns length MDS of each PMM2 variant model using Amber with the Amber forcefield[52,53]. Each PMM2 model was placed in a rectangular simulation cell with walls 12 Å from the nearest atom. The simulation cell was then filled with TIP3P waters at a density of 0.997 g/L and 150 mM Na$^+$/Cl$^-$ ions. The pH was set to 7.4, and amino acid protonation states were adjusted based on pKa. Long-range electrostatics were calculated using the Particle-Mesh Ewald (PME) Coulombic interaction method, with a periodic boundary condition cutoff of 7.86 Å. Pre-equilibration started with 3 stages of 10,000 timesteps (10 ps) each to energy to minimize the system using Steepest Descent (SD) Polak-Ribiere Conjugate Gradient (PRCG)[54], and gradually relaxing restraints. Two stages of 5 ps energy minimization commenced, followed by heating for 10,000 ps under soft restraints of 10 and 5 kcal/mol·Å$^2$, respectively, applied to all backbone atoms and metals. Next, 5 ps of energy minimization with solute restraints reduced to 1 kcal/mol·Å$^2$ initialized protein relaxation. The equilibrium phase was then completed by slowly heating the system from 80 to 310 K over the course of 4000 ps, while simultaneously relaxing protein atom restraints until all atoms were unrestrained. Finally, analytical MDS was carried out under the isothermal-isobaric (NPT) ensemble with a timestep of 2.5 fs, constraining X-H bonds with the SHAKE algorithm[55]. System pressure was enforced at 1 bar by readjustment of the simulation cell volume at a 1 ps relaxation time to maintain solvent density. Temperature was maintained by randomly assigning water velocities to fit a Boltzmann distribution. The analytical simulations consist of 10 ns of sampling (two frames archiving: start/finish for faster cpu speed). An identical protocol was followed for all 1454 variants, creating a total of ~15 microseconds of sampling.

**MDS feature extraction and analysis**
We performed quantitative analysis on a series of properties ("features") derived from each MDS to assess the conformational alterations induced by each mutation in the PMM2 protein. For each PMM2 missense variant, we performed 10 ns molecular dynamics (MD) simulations using the Desmond simulation engine in the Schrödinger Suite. By default, this setup generates 1001 trajectory frames, with snapshots saved every 10 picoseconds. For each frame in the trajectory, we computed a comprehensive set of structural and dynamic descriptors, including: RMSD, Rg, SASA, Free energy of stability, Hydrogen bond count, Amino acid contact density, Tensor of inertia, Secondary structure composition (α-helices, β-sheets, coils. Rather than using the full 10 ns blindly, we inspected the temporal profiles of these features across the trajectory and selected the region in which the features exhibited local convergence or stabilization, suggesting that the system had reached a relatively equilibrated state. The final values used as input features for our machine-learning models were the averages calculated over this stable portion of the trajectory. This approach allowed us to capture meaningful and representative dynamic behavior while minimizing the influence of initial structural fluctuations. The frequency profiles of key features throughout the simulation are shown in Fig. S6.

Rg was calculated using MDtraj[56] by determining the root-mean-square distance of the protein's atoms from their center of mass. Lower Rg values indicate a more compact structure, whereas higher values suggest a more expanded conformation. SASA was calculated with MDtraj using the Shrake-Rupley algorithm[57] to estimate the surface area accessible to solvent molecules. RMSD was calculated using MDtraj using the least squares fitting method to measure the average deviation of the protein's atomic positions from the WT PMM2 reference structure over time. Higher RMSD values suggest significant conformational changes. The tensor of inertia was calculated using MDtraj by evaluating the distribution of the protein's mass around its principal axes to understand its rotational properties. It provides information on the protein's shape and how it changes during the simulation. The Schrödinger Suite (Schrödinger Release 2024-1: Schrödinger, LLC, New York, NY, 2024) was utilized to calculate the free energy of stability via the MM-GBSA method[58]. This approach integrates molecular mechanics energies with solvation energies from the generalized Born model[59] and a surface area term. Lower free energy values correspond to more stable conformations. Amino acid contacts were determined by counting pairs within 0.5 Å of each other. Hydrogen bonds were identified based on geometric criteria involving donor-acceptor distance and angle cutoffs. Secondary structure elements were analyzed using the DSSP algorithm[60], which assigns classifications like α-helices and β-sheets based on hydrogen bonding patterns and geometric criteria.

**Labeling mutations by class**
We meticulously labeled the PMM2 mutations by thoroughly evaluating clinically recorded genetic sequencing results. ProteinPaint[61] in hg19 was used to identify all missense variants in *PMM2* (NM_000303) that have been recorded in ClinVar[38–40] to the date it was accessed (10/12/2023). The variants with clinical significance of pathogenic (P), likely pathogenic (LP), likely benign (LB), or benign (B) were recorded, resulting in 60 P and/or LP and 2 B and/or LB (Table S1).

Because only 2 true benign variants are not enough to train an AI algorithm, the gnomAD database[42,43] version 4 was accessed to elucidate more benign variants. Since all people whose sequenced genomes are collected in gnomAD are 18 years or older, and PMM2-CDG presents at a very early age with an autosomal recessive inheritance pattern, we can conclude that homozygous variants in this database are benign. Table S2 portrays data for 11 such homozygous variants that have an allele frequency indicative of recessivity. These variants were listed in ClinVar as B, LB, or VUS as indicated (Table S2), but we conclude that they are all true benign[1].

In collaboration with CIM in MCF and MCR, we identified patients carrying compound heterozygous pathogenic variants and collected extensive data on both the benign and damaging nature of PMM2 variants. Additionally, we obtained further damaging mutations for PMM2 from the Human Gene Mutation Database[41].

**Balancing the dataset for AI model training**
When training AI models with imbalanced data, such as our dataset of 54 benign, 97 damaging, and 6 ambiguous mutations (Fig. S4), it is crucial to use techniques that mitigate bias towards the majority class and enhance learning from the minority classes. The following approaches were employed to address class imbalance and the challenges associated with small datasets:

**Oversampling the minority class.** To increase the representation of the minority classes, we replicated existing samples of benign and ambiguous mutations. This method provides a higher representation of the minority class in the sample. This approach can lead to overfitting if the minority class instances are merely duplicated, but is, particularly useful for small datasets as it increases the amount of available training data.

**Downsampling the majority class.** To prevent the majority class (damaging mutations) from dominating the learning process, we reduced the number of damaging instances. While this approach helps balance the data and prevents the model from being biased towards the majority class, it may lead to the loss of important information. For small datasets, this method balances the data without generating additional data.

**Synthetic minority oversampling technique (SMOTE).** SMOTE was applied to generate synthetic samples for the minority classes by interpolating between existing minority class samples[44]. This technique reduces the risk of overfitting associated with simple oversampling and creates a more balanced dataset without losing majority class information. SMOTE is particularly effective for small datasets as it increases the size of the minority class without requiring additional real-world data. This method provided the best results for our dataset and was used when moving forward. This approach resulted in an equal number of samples across the benign, damaging, and ambiguous categories, as illustrated in Supplementary Fig. S5.

**Algorithmic adjustments.** We adjusted the algorithms by weighting the classes differently, giving higher importance to the minority classes during training. This approach helps prioritize the learning of minority class instances without resampling or generating synthetic data, but requires careful tuning of class weights to avoid introducing new biases. This method is beneficial for small datasets as it does not rely on data augmentation.

**Ensemble methods**
Bagging. This involves taking multiple random samples from the training dataset with replacement and training a model for each sample. The final prediction is averaged across all sub-models, reducing variance and improving model robustness against imbalanced data. Bagging maximizes the use of available data in small datasets.

Boosting. This involves training a sequence of models, where each model aims to correct the errors of its predecessor. We used the AdaBoost[62,63] technique to focus on hard-to-classify instances, aiming to convert a set of weak classifiers into a strong one. Boosting can significantly enhance model performance on imbalanced datasets by focusing on difficult instances.

With these balanced datasets by SMOTE (*Balanced_SMOTE_X_Features.xlsx* and *Balanced_SMOTE_Y_Labels.xlsx*) (https://doi.org/10.6084/m9.figshare.29132102), we trained a supervised multiclass classifier where each mutation was assigned a unique label: benign (0), damaging (1), or ambiguous (3). These categories are mutually exclusive, meaning each mutation belongs to only one class. An additional label, unknown (2), was reserved for mutations that could not be found in available human mutation databases and thus lacked any existing annotation; these were excluded from model training and evaluation, and instead preserved for benchmarking our model's predictions against other established mutation effect prediction programs (see *Raw_data.xlsx* available here: https://doi.org/10.6084/m9.figshare.29132102).

To address class imbalance, we applied SMOTE to the training set to ensure equal representation of the three labeled categories (0, 1, 3). During training, the model learned to classify each mutation into one of these three classes, without the need for explicitly defining "opposing" samples. For example, when learning to classify benign mutations, the model inherently distinguishes them from damaging and ambiguous mutations as part of the multiclass classification objective.

The model was evaluated on the unbalanced, original dataset, and classification performance was assessed using multiclass ROC curves and a confusion matrix, as shown in Fig. 4.

### Scaling the features for data processing

Scaling features is a crucial step in data preprocessing for machine-learning models because different features often have different units of measurement and ranges. In our dataset, this variability is evident. For example, the feature SASA ranges from approximately 130 to 133, while RMSD ranges from about 0.088 to 0.108, and the tensor of inertia ranges from about 24,275 to 27,757. These disparities can lead to issues when using models that rely on distance calculations, such as KNN, or gradient descent-based algorithms like linear regression, logistic regression, or DNN, as these algorithms might unduly weight the larger-scale features more, leading to biased and suboptimal model performance. To address this issue, we employed the StandardScaler() method from the scikit-learn library[64] for feature scaling.

### Data splitting and cross-validation

To ascertain the robustness of our AI models, we implemented a stratified partitioning of the dataset into distinct training and testing cohorts, supplemented by rigorous cross-validation procedures.

**Train-test split**. Utilizing the 'train_test_split' function from the scikit-learn library[64], we divided the dataset—previously balanced via SMOTE —into training and testing sets in an 80:20 ratio. The 'random_state' parameter was fixed at 42 to guarantee the reproducibility of our results.

**Cross-validation**. We adopted a k-fold cross-validation approach with 10 folds. The training dataset was partitioned into 10 equal-sized subsets. Each model was then trained on 9 of these subsets and validated against the remaining one. This cycle was iterated 10 times, with each subset serving once as the validation set. Additionally, a separate subset of mutations, which were not included in the SMOTE-balanced training set, was retained as an independent test set to rigorously assess the generalization capability of the trained model.

### Model training and optimization

We employed the following machine-learning algorithms as multiclass classification models to predict the nature of PMM2 mutations by distinguishing among three distinct categories: benign, damaging, and ambiguous. Here, we use the terms "VUS" and "ambiguous" interchangeably to refer to the same category of mutations. Models were run using TensorFlow[65] and carefully tuned to enhance performance and accuracy.

Benchmark models AlphaMissense[30], REVEL[9] and PROVEAN[14,15] were performed according to standard protocols.

Each model was trained using a SoftMax or equivalent probabilistic output function, enabling direct multiclass prediction rather than a series of binary classifications.

**RF**. The RF model is an ensemble method that constructs multiple DT during training and outputs the class that is the mode of the classes predicted by individual trees. This approach reduces overfitting, handles large datasets well, and provides insights into feature importance. We adjusted the number of trees (n_estimators = 100), maximum depth (max_depth = 7), and other hyperparameters to improve accuracy.

**DNN**. The DNN model, inspired by the human brain, consists of layers of interconnected nodes ("neurons") that learn to classify input data through backpropagation. This model is highly flexible and can capture complex relationships and interactions in the data. We experimented with different architectures, including the number of dense layers (up to 75) and neurons, activation functions (softmax), and learning rates to optimize performance (using the 'Adam' optimizer). The loss function used was 'categorical_crossentropy'.

**SSL**. SSL combines a small amount of labeled data with a large amount of unlabeled data during training, leveraging the unlabeled data to improve learning accuracy. This approach is particularly useful when labeled data is scarce, as in our case with the benign or ambiguous class. We used the 'LabelSpreading' algorithm with a radial base function (RBF) kernel, adjusting the parameters to 'kernel = 'rbf'', 'alpha = 0.2', and 'gamma = 20' to enhance learning and improve model performance.

**DT**. A DT model splits the data into subsets based on the value of input features, creating a tree-like structure of decisions. This model requires little data preprocessing. We tuned the maximum depth (max_depth = 5), minimum samples split (min_samples_split = 20), and other criteria to prevent overfitting and improve generalization.

**GBC**. The GBC model is an ensemble technique that builds multiple weak learners (typically DT) sequentially, with each new tree correcting errors made by the previous ones. This approach offers high accuracy, handles overfitting well, and provides feature importance. We adjusted the learning rate (learning_rate = 0.05), number of boosting stages (n_estimators = 200), and the maximum depth (max_depth = 4) of individual trees to enhance model performance.

**SVM-rbf**. The SVM-rbf model finds the optimal hyperplane to separate different classes by transforming the input data into a higher-dimensional space using the RBF kernel. This model is effective in high-dimensional spaces and works well with a clear margin of separation. We optimized the regularization parameter (C) and the RBF kernel parameter (gamma = 20) to improve classification accuracy.

**KNN**. The KNN method classifies a sample based on the majority class among its *k* nearest neighbors in the feature space. This non-parametric approach works well with small datasets and requires no training phase. We experimented with different values of *k* (n_neighbors = 4) and distance metrics to find the optimal configuration.

**LR**. LR is a linear model that uses a logistic function to model the probability of a binary or multiclass response based on input features. We adjusted the regularization strength (C) and utilized different solvers to find the best fit for the data.

### Feature importance calculation

To accurately assess the predictive significance of each feature in classifying PMM2 mutations, we employed a DT classifier and its associated feature importance metric. The classifier was trained on a normalized dataset, and feature importance values were derived using the classifier's "feature_importances" attribute. These values numerically represent each feature's contribution to the model's predictive accuracy. To ensure the stability and robustness of the feature rankings, we performed stratified 10-fold cross-validation. Stratified 10-fold cross-validation was chosen to maintain the proportion of each class in the training and validation sets. Specifically, the dataset was divided into 10 subsets, ensuring each subset contained approximately the same percentage of samples of each class as the original dataset. The model was then trained on 9 of these subsets and validated on the remaining subset. This process was repeated 10 times, with each subset used exactly once as the validation data.

**Creation of humanized wildtype animals**. The coding sequence for the canonical PMM2 (isoform 1) was extracted from the UniProt database (https://www.uniprot.org/). The sequence was codon-optimized for transgene expression in *C. elegans*[66]. Two synthetic introns were introduced, with optimization to ensure splicing specific to these introns. The sequence was introduced into the *C. elegans* ortholog F52B11.2 locus using CRISPR-Cas9 gene editing[67]. Two sgRNA target sites in F52B11.2 were selected, one in the first exon of F52B11.2 and the other in the last exon of F52B11.2. A long donor homology oligonucleotide (dhODN) was designed to have at least 35 base pairs of homology on the outside ends of the cut sites. The dhODN sequence containing two donor homology sequences flanking the codon-optimized PMM2 sequence was synthesized by Integrated DNA Technologies (IDT DNA). Care was taken to design for the elimination of the sgRNA target sites to avoid recutting of the edited locus. The dhODN was co-injected into the hermaphrodite gonad with the appropriate sgRNAs (Synthego), Cas9 (PNABio) and dpy-10 co-CRISPR reagents[68]. Animals were isolated based on the dpy-10 co-CRISPR phenotype and screened for PMM2 insertion by PCR. In animals homozygous for the edit, verification of the desired edit was confirmed by PCR followed by DNA sequencing. Transgenic lines were also confirmed to be wild-type at the dpy-10 co-CRISPR locus.

**Creation of variant animals**. For each PMM2 variant, a set of sgRNAs was selected to flank the site of interest. A donor homology oligonucleotide (dhODN) was designed to have at least 35 base pairs of homology on the outside ends of the cut sites. In the interval between the cut sites, the DNA was recoded with a new sequence, containing both the desired amino acid change and silent mutations that block recutting. The dhODN was co-injected into the hermaphrodite gonad with the appropriate sgRNAs and Cas9 and dpy-10 co-CRISPR reagents[68]. Animals displaying the dpy-10 co-CRISPR phenotype were selected, and High-Resolution Melt Analysis (HRMA) was used to identify those containing the edit of interest. Edits were verified by Sanger sequencing, and transgenic lines were confirmed to be wild-type at the dpy-10 co-CRISPR locus.

**Statistics and reproducibility**. We analyzed a total of 1454 PMM2 variants. Of these, 1297 were clinically uncharacterized (unknown), 54 were benign, 97 were damaging, and 6 were labeled as ambiguous or of uncertain significance. To address class imbalance, we applied the SMOTE algorithm to generate synthetic samples for the minority classes. Feature scaling was performed using the StandardScaler() function from the scikit-learn library to normalize disparities in scale across features such as solvent-accessible surface area (SASA: ~130–133), root-mean-square deviation (RMSD: ~0.088–0.108), and tensor of inertia (~24,275–27,757). Without scaling, these disparities could bias models based on distance or gradient descent, including KNN, logistic regression, and deep neural networks. All analyses were performed on the full dataset, and reproducibility was ensured through standardized pre-processing and consistent application of methods across all models.

**Reporting summary**
Further information on research design is available in the Nature Portfolio Reporting Summary linked to this article.

## Data availability
Supplementary Data 1 for the entire dataset, generated during and analyzed during the current study, which are available at the FigShare repository, which is accessible here: https://doi.org/10.6084/m9.figshare.29132102.

## Code availability
The custom code used in this study is available at GitHub and can be accessed via the following URL: https://github.com/thom-DEC/dynamicasome-CommBiol-2025. There are no restrictions on the availability of the code. The version of the code used in the analysis is 1, and any future updates will be managed through the same repository with version control.

## Acknowledgements
T.R.C. would like to thank the Center for Individualized Medicine for supporting this research initiative. T.R.C. would like to thank Dr. Klaas Wierenga for useful discussions on genomics and genetic pathogenicity and directions to database annotations (e.g., HMGD, ClinVar, etc).

## Author contributions
Conceptualization, T.R.C., N.N.I.; Methodology, N.N.I. and T.R.C.; Software, N.N.I. and T.R.C.; Validation, N.N.I., M.A.C., B.J., T.J.B. and T.R.C.; Formal analysis, N.N.I., M.A.C., J.M.F., and T.R.C.; Investigation, N.N.I., M.A.C., J.M.F., B.J. and T.R.C.; Resources, N.N.I. and T.R.C.; Data curation, C.W., N.N.I., M.A.C., J.M.F. and T.R.C.; Writing—original draft, T.R.C., N.N.I., J.M.F., R.C., B.J., T.J.B., C.T. and M.A.C.; Writing—review and editing, N.N.I., R.C., B.J., T.J.B. and T.R.C.; Visualization, N.N.I., J.M.F., and M.A.C.; Supervision, T.R.C.; Project administration, T.R.C.; Funding acquisition, T.R.C. All authors have read and agreed to the published version of the manuscript.

## Competing interests
The authors declare no competing interests.

## Additional information
**Supplementary information** The online version contains supplementary material available at https://doi.org/10.1038/s42003-025-08334-y.

**Correspondence** and requests for materials should be addressed to Thomas R. Caulfield.

**Peer review information** *Communications Biology* thanks Yaning Yang and the other, anonymous, reviewer(s) for their contribution to the peer review of this work. Primary handling editor: Aylin Bircan.

**Reprints and permissions information** is available at http://www.nature.com/reprints

**Publisher's note** Springer Nature remains neutral with regard to jurisdictional claims in published maps and institutional affiliations.